%% file: main.tex
\documentclass{article}

\usepackage[final]{neurips_2020}

\usepackage[utf8]{inputenc} 
\usepackage[T1]{fontenc}    
\usepackage{hyperref}       
\usepackage{url}            
\usepackage{booktabs}       
\usepackage{amsfonts}       
\usepackage{nicefrac}       
\usepackage{microtype}      

\input{macros}
\title{Social Choice with Changing Preferences: Representation Theorems and Long-Run Policies}

\author{
  Kshitij Kulkarni \\
  Department of EECS\\
  University of California, Berkeley\\
  Berkeley, CA 94709 \\
  \texttt{ksk@eecs.berkeley.edu} \\
  \And
  Sven Neth \\
  Department of Philosophy \\ 
  University of California, Berkeley \\
  Berkeley, CA 94709 \\
  \texttt{nethsven@berkeley.edu} \\
}

\begin{document}

\maketitle

\begin{abstract}
We study group decision making with changing preferences as a Markov Decision Process. We are motivated by the increasing prevalence of automated decision-making systems when making choices for groups of people over time. Our main contribution is to show how classic representation theorems from social choice theory can be adapted to characterize optimal policies in this dynamic setting. We provide an axiomatic characterization of MDP reward functions that agree with the Utilitarianism social welfare functionals of social choice theory. We also provide discussion of cases when the implementation of social choice-theoretic axioms may fail to lead to long-run optimal outcomes. 
\end{abstract}

\section{Introduction}
Social choice theory \citep{arrow1951} is a classic subfield of economics and philosophy that seeks to identify decisions that a social planner may make for a group based on the preferences of the group's members. In particular, the standard theorems of social choice are so-called \textit{representation theorems} that provide constraints on the kinds of social alternatives that may be chosen given axioms on individual preferences. However, social choice generally operates in an environment where preferences are static and are not shaped by previous decisions that have been made by the  social planner \citep{Gaertner2006}. This leads to various critiques of social choice, for example from Pettigrew, who argues that utility functions of individuals \textit{change over time} \citep{Pettigrew2020}, and Parkes and Procaccia, who study the problem of group decision making with changing preferences from a voting-theoretic perspective \citep{ParkesProcaccia2014}. In recent years, there has also been work on studying dynamic preferences in other contexts \citep{VelosoDynamic2005} \citep{Tadepalli2005DynamicPreferences}. 

In parallel, there is a standard theory of dynamic decision-making in the Markov Decision Process (MDP) literature \citep{Puterman1994} that studies memoryless state-transition models with reward functions and policies that maximize long-run rewards. MDPs are used in many decision-making tasks, most commonly in reinforcement learning applications \citep{Barto89learningAS}.

As automated decision-making systems begin to make a larger fraction of choices for groups of individuals, we believe that a theory which joins the rigorous representation theorems of social choice theory with the dynamic nature of sequential decision-making is required. We aim to provide an optimality criterion for policies in such a setting by drawing on the rich existing work in social choice theory.

\subsection{Our Results and Contributions}
Our main contribution is to use a particular model of dynamic social choice, called a \emph{Social Choice MDP} \citep{ParkesProcaccia2014}, and to show how we can draw on representation theorems to constrain reward functions and optimal policies to agree with the social welfare functional of Utilitarianism. In order to prove this representation theorem, we connect the class of reward functions of the Social Choice MDP with the individual utility functions of the agents in the group by providing axiomatic constraints which are necessary and sufficient for the reward function of the Social Choice MDP to agree with Utilitarianism. We also characterize the class of policies that arise as a result of implementing the long-run maximization of (Quasi-)Utilitarian rewards, and show that these policies lead to reasonable optimality criteria. This leads to axiomatic constraints on the value function of the optimization problem. Finally, we note that there are axioms that are standard in social choice theory, but whose validity breaks down in the dynamic setting. The most prominent of these is the (local) version of the Pareto axiom \citep{ParkesProcaccia2014}. 
\subsection{Previous Work}

There has been previous work on studying dynamic social choice with evolving preferences by Parkes and Procaccia \citep{ParkesProcaccia2014}. The main difference between this previous work and our contribution is that we use a different way to map social choice concepts to MDPs. While Parkes and Procaccia focus on axioms on social choice functions as constraints on policies, we focus on axioms on social welfare functionals as  constraints on the reward function. The significance of this difference is that we can draw on the rich work in social choice theory on representation theorems for social welfare functionals, especially representation theorems for Utilitarianism \citep{bentham1789}. Relatedly, while Parkes and Procaccia assume that group members have only ordinal preferences, we assume that their preferences are represented by (cardinal) utility functions.

\section{Social Choice Theory} 

In this section, we introduce the basics of (classic) social choice theory \citep{arrow1951, Gaertner2006}. We start with a non-empty, finite set $V$ of \emph{group members} and a non-empty, finite set $X$ of \emph{social alternatives}. Let $\mathcal{U}(X)$ be $\{ u \mid u : X \to \mathbb{R} \}$, the set of all utility functions over the social alternatives. Then, we define:
\begin{definition}
A \emph{profile} is a function $U : V \to \mathcal{U}(X).$
\end{definition}
This means that a profile is an assignment of utility functions to group members. For every $i \in V$, we write $U_i(x)$ as shorthand for $U(i)(x)$. Let $\mathscr{U}$ be the set of all profiles.

In social choice theory, we are interested in how a group, or a `social planner', should make decisions based on the preferences of all group members. There are different ways of formalizing this question. First, we can study social choice functions:
\begin{definition}
A \emph{social choice function (SCF)} is a map $f : \mathcal{D} \to X$, where $\mathcal{D}$ is some set of profiles.
\end{definition}
Given some profile $U \in dom(f)$, a SCF $f$ selects a preferred social alternative $f(U) \in X$. Note that a SCF $f$ must select a unique $x \in X$ for each $U \in \mathcal{D}$. This is a potential drawback of SCFs, as there may be situations in which different alternatives are equally good. In this case, SCFs require the introduction of arbitrary tie-breakers. Further, SCFs do not encode any information about the ranking among the social alternatives which are not chosen. We can avoid both of these problems by focusing instead on social welfare functions:
\begin{definition}
A \emph{social welfare functional (SWF)} is a map $f : \mathcal{D} \to \mathcal{B}(X)$, where $\mathcal{D}$ is some set of profiles and $\mathcal{B}(X)$ is the set of all binary relations on $X$.
\end{definition}
Given some profile $U \in dom(f)$, a SWF $f$ returns a binary relation on $X$, which we interpret as a `social preference relation'. For any profile $U$, we write $xf(U)y$ if $(x,y) \in f(U)$. The intended interpretation of $xf(U)y$ is that `$x$ is socially preferred to $y$'. We write $xP(f(U))y$ if $xf(U)y$ and not $yf(U)x$. The intended interpretation of $xP(f(U))y$ is that `$x$ is strictly socially preferred to $y$'. 

Much work in social choice theory focuses on axioms which are imposed either on SCFs or SWFs, for example some form of the Pareto principle. Here are Pareto axioms for SCFs and SWFs:
\begin{quote}
SCF $f$ satisfies \textbf{Pareto (SCF)} if for all $U \in dom(f)$ and all $x,y \in X$, if $U_i(x) > U_i(y)$ for all $i \in V$, then $f(U) \not = y$ .
\end{quote}
\begin{quote}
SWF $f$ satisfies \textbf{Pareto (SWF)} if for all $U \in dom(f)$ and all $x,y \in X$, if $U_i(x) > U_i(y)$ for all $i \in V$, then $xP(f(U))y$.
\end{quote}
Research in social choice theory focuses in particular on \emph{representation theorems}: finding a set of axioms which are necessary and sufficient for a SCF or SWF to be representable by a certain functional form \citep{blackorby2002}. An example is the SWF of \emph{Utilitarianism}:
\begin{definition}
SWF $f$ is \emph{Utilitarianism} if for all $U \in dom(f)$, $x,y \in X$, 
\begin{equation*}
xf(U)y \iff \sum_{i \in V}U_i(x) \geq \sum_{i \in V} U_i(y).     
\end{equation*}
\end{definition}

Consider the following axioms:
\begin{quote}
SWF $f$ satisfies \textbf{Universal Domain} if $dom(f)$ is the set of all profiles.    
\end{quote}

\begin{quote}
SWF $f$ satisfies \textbf{Transitivity} (\textbf{Completeness}) if for all $U \in dom(f)$, $f(U)$ is transitive (complete).    
\end{quote}

\begin{quote}
SWF $f$ satisfies \textbf{Independence of Irrelevant Alternatives (IIA)} if for all $U, U' \in dom(f)$ and $x,y \in X$, if $U_i(x) = U'_i(x)$ and $U_i(y) = U'_i(y)$ for all $i \in V$, then $xf(U)y$ if and only if $xf(U')y$. 
\end{quote}

\begin{definition}
Two profiles $U$ and $U'$ satisfy \emph{cardinal unit comparability}, written $U \sim_{CUC} U'$, if there is a $\beta \in \mathbb{R}$ with $\beta > 0$ and for every $i \in V$, there is some $\alpha_i \in \mathbb{R}$ such that for all $x \in X$, $U_i(x) = \alpha_i + \beta U'_i(x)$.
\end{definition}

\begin{quote}
SWF $f$ satisfies \textbf{CUC-Invariance} if for all $U,U' \in dom(f)$, if $U \sim_{CUC} U'$, then $f(U) = f(U')$.     
\end{quote}

\begin{quote}
SWF $f$ satisfies \textbf{Functional Anonymity} if for all $U,U' \in dom(f)$ and permutations $\rho : V \to V$, if for all $i \in V$, $U'_i = U_{\rho(i)}$, then $f(U) = f(U')$.     
\end{quote}

It has been shown that these axioms characterize Utilitarianism:

\begin{theorem}\label{thm1}
A SWF $f$ is Utilitarianism if and only if $f$ satisfies \textbf{Universal Domain}, \textbf{Transitivity}, \textbf{Completeness}, \textbf{IIA}, \textbf{Pareto (SWF)}, \textbf{CUC-Invariance} and \textbf{F-Anonymity} \citep{Aspremont1977}.
\end{theorem}

\section{Dynamic Decision-Making for Groups}
Social choice theory normally considers static decision-making for groups. While this is amenable to analysis (through representation theorems), there is a critique of social choice in that it does not consider the case when preferences shift over time. Here, we consider the dynamic setting where the group members' preferences are changing over time according to a probabilistic model that is known to the social planner. 

\subsection{Markov Decision Processes}
This section introduces our model for dynamic decision-making for groups. We consider Markov Decision Processes, which are memoryless state-transition models along with a reward function, which we define as:

\begin{definition}
A \emph{Markov Decision Process} is a tuple $\langle \mathcal{S}, \mathcal{A}, R, P\rangle$ where $\mathcal{S}$ is a finite non-empty set, $\mathcal{A}$ is a finite non-empty set, $P: \mathcal{S} \times \mathcal{A} \to \mathbb{R}$ is a probability function and $R : \mathcal{S} \times \mathcal{A} \to \mathbb{R}$ is a function.
\end{definition}
Further, the probability function satisfies the \emph{Markov assumption}, which means that the probability of the next state only depends on the current state-action pair. We also define the notion of a policy: 
\begin{definition}
A (deterministic) \emph{policy} is a function $\pi : \mathcal{S} \to \mathcal{A}$.
\end{definition}

\subsection{The Social Choice MDP Model}
Given the static models from social choice and the dynamic, state-transition based models from the MDP literature, we seek to define a model for dynamic decision-making when group members' preferences are shifting over time in response to actions taken by the social planner. 

\begin{definition}
A \emph{Social Choice Markov Decision Process} is a tuple $\langle \mathcal{S}, \mathcal{A}, R, P\rangle$ where $\mathcal{S}$ is a non-empty finite set of profiles $U : V \to \mathcal{U}(X)$, $\mathcal{A}$ is the set of finite social alternatives $X$, $P: \mathcal{S} \times \mathcal{A} \to \mathbb{R}$ is a probability function and $R : \mathscr{U} \times \mathcal{A} \to \mathbb{R}$ is a function.
\end{definition}
Our model differs from other Social Choice MDP models in two ways: first, we assume cardinal preferences, which give rise to MDP state spaces that are comprised of assignments of utility functions to group members, and second, our reward functions are defined over $\mathscr{U}$,  the set of all profiles, in order for our representation theorems to hold. 

\subsection{From SCFs to SWFs}

Note that in a Social Choice MDP, a policy $\pi: \mathcal{S} \to \mathcal{A}$ is a social choice function $f : \mathcal{D} \to X$, since $\mathcal{S}$ is a set of profiles and $\mathcal{A}$ is a set of social alternatives. Thus, \citep{ParkesProcaccia2014} apply insights from the social choice literature to characterize policies in Social Choice MDPs. However, observe that there is also a correspondence between reward functions and SWFs. In particular, every reward function $R$ in a Social Choice MDP induces a social welfare functional $f_R: \mathcal{D} \to \mathcal{B}(X)$. 
\begin{definition}
Given a reward function $R$, we define the corresponding SWF $f_R$ for every profile $U$ and all $x,y \in X$:
\begin{equation*}
x f_{R}(U) y \iff R(U,x) \geq R(U,y).
\end{equation*}
\end{definition}
We can use this correspondence to use social choice axioms on SWFs as constraints on the reward function. One natural choice is the Utilitarian reward function: for every $U \in \mathcal{S}$ and $a \in \mathcal{A}$, $R(U,a) = \sum_{i \in V} U_i(a)$. However, instead of requiring the reward function to be strictly Utilitarian, we can also focus on the weaker requirement that it must agree with Utilitarianism up to strictly increasing transformations:
\begin{definition}
A reward function $R : \mathcal{S} \times \mathcal{A} \to \mathbb{R}$ is \emph{Quasi-Utilitarian} if for every $U \in \mathcal{S}$ and $a \in \mathcal{A}$, $R(U,a) = f \left(\sum_{i \in V} U_i(a) \right)$, where $f: \mathbb{R} \to \mathbb{R}$ is a strictly increasing function.
\end{definition}

\section{Quasi-Utilitarian Characterization}
In this section, we introduce constraints on the reward function $R$ which entail that $R$ agrees with the Utilitarianism social welfare functional and give a characterization of the policies generated by these Quasi-Utilitarian reward function.

\subsection{Reward functions and SWFs}

Using the mapping from reward functions and SWFs introduced above, we will impose the following axioms on the reward function $R$, which correspond to axioms on the induced SWF $f_R$:
\begin{quote}
Reward function $R$ satisfies \textbf{Pareto (SWF)} if for all $U \in dom(f_R)$ and all $x,y \in X$, if $U_i(x) > U_i(y)$ for all $i \in V$, then $xP(f_R(U))y$.
\end{quote}

\begin{quote}
Reward function $R$ satisfies \textbf{Independence of Irrelevant Alternatives (IIA)} if for all $U, U' \in dom(f_R)$ and $x,y \in X$, if $U_i(x) = U'_i(x)$ and $U_i(y) = U'_i(y)$ for all $i \in V$, then $xf_R(U)y$ if and only if $xf_R(U')y$.    
\end{quote}

\begin{quote}
Reward function $R$ satisfies \textbf{CUC-Invariance} if for all $U,U' \in dom(f_R)$, if $U \sim_{CUC} U'$, then $f_R(U) = f_R(U')$.     
\end{quote}

\begin{quote}
Reward function $R$ satisfies \textbf{Functional Anonymity} if for all $U,U' \in dom(f_R)$ and permutations $\rho : V \to V$, if for all $i \in V$, $U'_i = U_{\rho(i)}$, then $f_R(U) = f_R(U')$.     
\end{quote}
Then, we can show:
\begin{theorem}\label{thm2}
The following are equivalent for any Social Choice MDP:
\begin{enumerate}
    
    \item $R$ satisfies \textbf{Pareto (SWF)}, \textbf{IIA}, \textbf{CUC-Invariance} and \textbf{Functional Anonymity}.
    
    \item $R$ agrees with Utilitarianism, so for any profile $U$ and $x,y \in X$, we have
    \begin{equation*}
    R(U,x) \geq R(U,y) \iff \sum_{i \in V} U_i(x) \geq \sum_{i \in V} U_i(y).    
    \end{equation*}
    Equivalently, $R$ is Quasi-utilitarian. 
    
\end{enumerate}
\end{theorem}

\subsection{Long Run Maximization}

To get from reward function to optimal policies, we need to make additional assumptions. In this section, we draw on standard results from the MDP literature to argue for a particular kind of policy.
\begin{definition}
A \emph{value function} is a map $V : \Pi \times S \to \mathbb{R}$, where $\Pi$ is the set of all policies and $S$ is the set of all states.
\end{definition}
Intuitively, $V(\pi, s)$ is the value of executing policy $\pi$ starting in state $s$.  Given a value function, we define:
\begin{definition}
The policy $\pi^*$ is \emph{optimal relative to $V$} if for all states $s \in S$, $\pi^* \in \underset{\pi \in \Pi}{\mathrm{arg\max\ }} V(\pi, s)$.
\end{definition}
We assume that the value function satisfies the \emph{Bellman equation} \citep{Puterman1994} for any $\pi \in \Pi$ and $s \in S$:
\begin{equation*}
V(\pi, s) = R(s, \pi(s)) + \gamma \sum_{s' \in S} p(s' \mid s, \pi(s)) V(\pi, s'),
\end{equation*}
 where $0 < \gamma < 1$. This means that the value of executing policy $\pi$ starting in state $s$ is the sum of the immediate reward $R(s, \pi(s))$ and the expected future value of executing $\pi$ in the next state, discounted by $\gamma$.\footnote{There are interesting questions about how to choose the discount rate which we cannot discuss here in detail, see e.g. \citep{FleurbaeyZuber2013}.} Now we can appeal to a standard result in the theory of MDPs \cite{sutton2018, szepesvari2010}:

\begin{theorem}\label{thm3}
Let $V : \Pi \times S \to \mathbb{R}$ be a value function. Then the following are equivalent for any MDP:
\begin{enumerate}

\item $V$ satisfies the Bellman equation.

\item $V$ is the expected sum of discounted future rewards. So, for any $\pi$ and $s$,
\begin{equation*}
V(\pi, s) = \mathbb{E} \left[ \sum_{t = 1}^{\infty} \gamma^{t} R(s_t, \pi(s_t)) \right],
\end{equation*}
where $s_t$ is a random variable describing the state after $t$ steps starting in state $s$ with policy $\pi$ and the expectation is taken relative to the transition model $P$.
\end{enumerate}
\end{theorem}

Taken together with theorem \ref{thm2}, we can use this result to characterize what we call the class of \emph{long-run quasi-utilitarian} policies:
\begin{definition}
Given a Social choice MDP, a policy $\pi^*$ is \emph{long-run quasi-utilitarian} if for all $s \in \mathcal{S}$,
\begin{equation*}
\pi^* \in \underset{\pi \in \Pi}{\mathrm{arg\max\ }} \mathbb{E} \left[\sum_{t = 1}^{\infty} \gamma^{t} f \left( \sum_{i \in V} U^{t}_{i}(a) \right) \right],
\end{equation*}
where $U^{t}$ is a random variable describing the profile after $t$ steps starting in state $s$ with policy $\pi$, the expectation is taken relative to the transition model $P$, and $ f: \mathbb{R} \to \mathbb{R}$ is strictly increasing.
\end{definition}
We propose this as a reasonable optimality criterion for group decision making under changing preferences. We can characterize this class as follows:

\begin{theorem}\label{thm4}
Given a Social choice MDP, assume that $V$ satisfies \textbf{Bellman equation} and $R$ satisfies \textbf{Weak Pareto}, \textbf{IIA}, \textbf{CUC-Invariance} and \textbf{Functional Anonymity}. Then, the following are equivalent for any policy $\pi$:
\begin{enumerate}
    \item $\pi$ is optimal relative to $V$,
    \item $\pi$ is long-run quasi-utilitarian.
\end{enumerate}
\end{theorem}

\section{Discussion}

We finish by discussing some consequences of our approach to group decision making with changing preferences. As noted above, there are two different ways of mapping social choice concepts to MDPs. First, we can think of policies as social choice functions (SCFs) and use axioms on SCFs to constrain policies. Second, we can exploit a correspondence between reward functions and social welfare functionals (SWFs), which is our distinctive contribution. We also noted earlier that there are two versions of the Pareto axiom for SCFs and SWFs respectively. The axioms for group decision making with changing preferences we defend here imply that our reward function satisfies the Pareto axiom for the SWF induced by the reward function:
\begin{quote}
Reward function $R$ satisfies \textbf{Pareto (SWF)} if for all $U \in dom(f_R)$ and all $x,y \in X$, if $U_i(x) > U_i(y)$ for all $i \in V$, then $xP(f_R(U))y$.
\end{quote}
However, the policies which satisfy our optimality criterion will \emph{not}, in general, satisfy the Pareto axiom for SCFs:
\begin{quote}
SCF $f$ satisfies \textbf{Pareto (SCF)} if for all $U \in dom(f)$ and all $x,y \in X$, if $U_i(x) > U_i(y)$ for all $i \in V$, then $f(U) \not = y$ .
\end{quote}
Applied to policies $\pi$, this axiom states that for all profiles $U$ and all $x,y \in X$, if $U_i(x) > U_i(y)$ for all $i \in V$, then $\pi(U) \not = y$. This means that if every group member assigns higher utility to social alternative $x$ than to social alternative $y$, $y$ will not be chosen by our policy. However, this will not be true in general. Suppose, for example, that $y$ leads, with high probability, to a future trajectory of high reward, while $x$ leads, with high probability, to a future trajectory of low reward. Then, a long-run optimal policy will often choose $y$ over $x$ even though all group members assign higher utility to $x$.

This is interesting, because Parkes and Proccacia seem to suggest that the latter version of the Pareto axiom is a normatively sound constraint on group decision making:
\begin{quote}
In the case of Pareto optimality, if at any point the members all prefer one choice to another then the latter choice should not be made by the organization. \citep{ParkesProcaccia2014}
\end{quote}
In our view, while Pareto optimality in this sense might perhaps be a compelling axiom in some social choice contexts, such as sequential voting, it is \emph{not} compelling in the context of long-run welfare optimization. This shows that once we focus on a dynamic setting with changing preferences, some of the traditional axioms of social choice theory lose their justification. Thus, it is important to study  group decision making with changing preferences on its own right.

\section*{Broader Impact}

Our work has potentially broad societal impact as automated decision systems become more ubiquitous, and the question of what constitutes optimality for such systems becomes very significant. We hope that by studying normative criteria for long-run optimality for group decision systems, we can contribute to making this impact positive. We also note that there is a large literature on social choice for welfare functionals that are not Utilitarianism, and in particular, welfare functionals that explicitly account for distributional properties like equity (ex. Leximin or Maximin \citep{blackorby2002}). We plan to consider these alternative approaches in future work.

\section*{Acknowledgements}
Kshitij Kulkarni is supported by CNS-1239166 from the National Science Foundation. Sven Neth is supported by the 2020 Global Priorities Fellowship by the Forethought Foundation.
\newpage
\bibliography{main}
\newpage
\section*{Appendix}

We begin by proving theorem \ref{thm2}, which adapts techniques from analogous results in the social choice literature (i.e. theorem \ref{thm1}):

\begin{proof}
First, we show that for any reward function $R$ in a Social Choice MDP, $f_R$ satisfies Universal Domain, Completeness and Transitivity. Consider an arbitrary profile $U \in \mathscr{U}$. We have $x f_R(U) y \iff R(U,x) \geq R(U,y)$, which is well defined since the domain of $R$ is $\mathscr{U} \times X$. Therefore, $f_R$ satisfies Universal Domain. Consider an arbitrary profile $U \in dom(f_R)$. By completeness of $\geq$ on $\mathbb{R}$, we have $R(U,x) \geq R(U,y)$ or $R(U,y) \geq R(U,x)$, so $x f_R(U) y$ or $y f_R(U) x$, so $f_R(U)$ is complete. Now assume $x f_R(U) y$ and $y f_R(U) z$ for some $x,y,z \in X$. It follows that $R(U,x) \geq R(U,y)$ and $R(U,y) \geq R(U,z)$. Therefore, $R(U,x) \geq R(U,z)$, so $x f_R(U) z$. Therefore, $f_R(U)$ is transitive. Since $U$ was arbitrary, $f_R$ satisfies Completeness and Transitivity.

Assume, in addition, that $R$ satisfies \textbf{Weak Pareto}, \textbf{IIA}, \textbf{CUC-Invariance} and \textbf{Functional Anonymity}. Therefore, $f_R$ satisfies \textbf{Pareto (SWF)}, \textbf{IIA}, \textbf{CUC-Invariance} and \textbf{Functional Anonymity}. So by theorem \ref{thm1}, $f_R$ is Utilitarianism:
\begin{equation*}
xf_R(U)y \iff \sum_{i \in V}U_i(x) \geq \sum_{i \in V} U_i(y).     
\end{equation*}
By definition of $f_R$, we have $x f_R(U) y \iff R(U,x) \geq R(U,y)$, so
\begin{equation*}
R(U,x) \geq R(U,y) \iff \sum_{i \in V}U_i(x) \geq \sum_{i \in V} U_i(y),     
\end{equation*}
so $R$ agrees with Utilitarianism.

We now prove the converse direction of the equivalence. That is, we want to show that if $R$ agrees with Utilitarianism, that is, for any profile $U$ and $x,y \in X$, we have 
\begin{equation}
    R(U,x) \geq R(U,y) \iff \sum_{i \in V} U_i(x) \geq \sum_{i \in V} U_i(y) 
\end{equation}
then $R$ satisfies \textbf{Pareto (SWF)}, \textbf{IIA}, \textbf{CUC-Invariance}, and \textbf{Functional Anonymity}. 

We start by showing that $R$ satisfies \textbf{Pareto (SWF)}. That is, we want to show that for all $U \in dom(f_R)$ and all $x,y \in X$, if $U_i(x) > U_i(y)$ for all $i \in V$, then $xP(f_R(U))y$. Assume that $U_i(x) > U_i(y)$ for all $i \in V$. Then, we know that $\sum_{i \in V} U_i(x) > \sum_{i \in V} U_i(y)$. Because $R$ agrees with Utilitarianism, then we know that $R(U,x) > R(U,y)$, and in turn, this means that the social welfare functional $f_R$ induced by $R$ satisfies $xP(f_R(U)) y$ for all $x,y \in X$, which is what we wanted to show. 

Next, we consider \textbf{IIA}. That is, we want to show that for all $U,U' \in dom(f_R)$ and $x,y \in X$, if $U_i(x) = U_i'(x)$ and $U_i(y) = U_i'(y)$ for all $i\in V$, then $x f_R(U)y$ if and only if $x f_R(U') y$. Assume that $U_i(x) = U_i'(x)$ and $U_i(y) = U_i'(y)$ and $x f_R(U) y$. Then, because $f_R$ is the SWF induced by $R$, we know that $R(U, x) \geq R(U,y)$, and furthermore, because $R$ agrees with Utilitarianism, we know that $\sum_{i \in V} U_i(x) \geq \sum_{i \in V} U_i(y)$. However, by the property that $U_i(x) = U_i'(x)$ and $U_i(y) = U_i'(y)$ for all $i \in V$, we get the inequality $\sum_{i \in V} U_i'(x) \geq \sum_{i \in V} U_i'(y)$. Therefore, once again, because $R$ agrees with Utilitarianism, we have $R(U', x) \geq R(U',y)$, and thus $x f_R(U')y$. The steps in this proof are reversible, and thus the converse direction follows as well. 

Now consider \textbf{CUC-Invariance}. For all $U,U' \in dom(f)$, we want to show that if $U \sim_{CUC} U'$, then $f_R(U) = f_R(U')$. Assume $U \sim_{CUC} U'$. By definition, there is a $\beta \in \mathbb{R}$ with $\beta > 0$ and for every $i \in V$, there is some $\alpha_i \in \mathbb{R}$ such that for all $x \in X$, $U_i(x) = \alpha_i + \beta U'_i(x)$. We have, for all $x,y \in X$,
\begin{equation*}
x f_R(U) y \iff \sum_{i \in V} U_i(x) \geq \sum_{i \in V} U_i(y) 
\end{equation*}
by assumption. By standard properties of summation, 
\begin{equation*}
 \sum_{i \in V} U_i(x) \geq \sum_{i \in V} U_i(y) \iff \sum_{i \in V} \beta U_i(x) \geq \sum_{i \in V} \beta U_i(y) \iff  \sum_{i \in V} \alpha_i + \beta U_i(x) \geq \sum_{i \in V} \alpha_i + \beta U_i(y)    
\end{equation*}
and by definition
\begin{equation*}
\sum_{i \in V} \alpha_i + \beta U_i(x) \geq \sum_{i \in V} \alpha_i + \beta U_i(y)  \iff   \sum_{i \in V} U'_i(x) \geq \sum_{i \in V} U'_i(y) \iff x f_R(U') y.
\end{equation*}
Therefore, $x f_R(U) y \iff x f_R(U') y$, so $f_R(U) = f_R(U')$.

We finish by showing \textbf{Functional Anonymity}. We want to show that for all $U,U' \in dom(f_R)$ and permutations $\rho : V \to V$, if for all $i \in V$, $U'_i = U_{\rho(i)}$, then $f_R(U) = f_R(U')$. Consider profiles $U$ and $U'$ and a permutation $\rho : V \to V$ such that for all $i \in V$, $U'_i = U_{\rho(i)}$. Now, for all $x,y \in X$:      
\begin{equation*}
x f(U) y \iff \sum_{i \in V} U_i(x) \geq \sum_{i \in V} U_i(y) 
\end{equation*}
Permutations do not affect the sum, so we have, for any permutation $\rho : V \to V$,
\begin{equation*}
\sum_{i \in V} U_i(x) \geq \sum_{i \in V} U_i(y) \iff \sum_{i \in V} U_{\rho(i)}(x) \geq \sum_{i \in V} U_{\rho(i)}(y),
\end{equation*}
and by definition
\begin{equation*}
\sum_{i \in V} U_{\rho(i)}(x) \geq \sum_{i \in V} U_{\rho(i)}(y) \iff \sum_{i \in V} U'_i(x) \geq \sum_{i \in V} U'_i(y) \iff x f_R(U') y,
\end{equation*}
which completes our proof.
\end{proof}

We proceed by proving theorem \ref{thm4}:

\begin{proof}
Consider a social choice MDP where $V$ satisfies the Bellman equation and $R$ satisfies \textbf{Weak Pareto}, \textbf{IIA}, \textbf{CUC-Invariance} and \textbf{Functional Anonymity}.

Assume that $\pi$ is optimal relative to $V$. Therefore, for all $s \in \mathcal{S}$
\begin{equation*}
\pi \in \underset{\pi \in \Pi}{\mathrm{arg\max\ }} V(\pi, s).
\end{equation*}

Since $V$ satisfies the Bellman equation, we have $V(\pi, s) = \mathbb{E} \left[ \sum_{t = 1}^{\infty} \gamma^{t} R(s_t, \pi(s_t)) \right]$ for all $s \in \mathcal{S}$ by theorem \ref{thm3}, so
\begin{equation*}
\pi^* \in \underset{\pi \in \Pi}{\mathrm{arg\max\ }} \mathbb{E} \left[\sum_{t = 1}^{\infty} \gamma^{t} R(s_t, \pi(s_t)) \right].
\end{equation*}
for all $s \in \mathcal{S}$.

By \textbf{Weak Pareto}, \textbf{IIA}, \textbf{CUC-Invariance} and \textbf{Functional Anonymity} and theorem \ref{thm2}, $R$ is quasi-utilitarian, so we have $R(U,a) = f \left(\sum_{i \in V} U_i(a) \right),$ where $f: \mathbb{R} \to \mathbb{R}$ is a strictly increasing function. Therefore,
\begin{equation*}
\pi^* \in \underset{\pi \in \Pi}{\mathrm{arg\max\ }} \mathbb{E} \left[\sum_{t = 1}^{\infty} \gamma^{t} f \left( \sum_{i \in V} U^{t}_{i}(a) \right) \right],
\end{equation*}
where $f$ is strictly increasing, so $\pi$ is long-run quasi utilitarian.

Assume that $\pi$ is long-run quasi utilitarian. By definition, for all $s \in \mathcal{S}$
\begin{equation*}
\pi^* \in \underset{\pi \in \Pi}{\mathrm{arg\max\ }} \mathbb{E} \left[\sum_{t = 1}^{\infty} \gamma^{t} f \left( \sum_{i \in V} U^{t}_{i}(a) \right) \right],
\end{equation*}
so
\begin{equation*}
\pi^* \in \underset{\pi \in \Pi}{\mathrm{arg\max\ }} \mathbb{E} \left[\sum_{t = 1}^{\infty} \gamma^{t}R(s_t, \pi(s_t)) \right],
\end{equation*}
where $R(U,a) = f \left( \sum_{i \in V} U^{t}_{i}(a) \right)$ for some strictly increasing $f$.
By theorem \ref{thm3}, since $V$ satisfies the Bellman equation, we have $V(\pi, s) = \mathbb{E} \left[ \sum_{t = 1}^{\infty} \gamma^{t} R(s_t, \pi(s_t)) \right]$ for all $s \in \mathcal{S}$.

Therefore, we have for all $s \in \mathcal{S}$
\begin{equation*}
\pi \in \underset{\pi \in \Pi}{\mathrm{arg\max\ }} V(\pi, s),
\end{equation*}
so $\pi$ is optimal relative to $V$.
\end{proof}
\end{document}

%% file: macros.tex
\usepackage{amsthm}
\newtheorem{definition}{Definition}
\newtheorem{theorem}{Theorem}

\usepackage{amssymb}
\usepackage{amsmath}
\usepackage{mathrsfs}  